\documentclass[fleqn,10pt]{wlscirep}
\usepackage[utf8]{inputenc}
\usepackage[T1]{fontenc}
\title{All-optical spin switching probability in [Tb/Co] multilayers}

\author[1,*,+]{L. Avil\'es-F\'elix}
\author[1]{L. Farcis}
\author[1]{Z. Jin}
\author[1]{L. \'Alvaro-G\'omez}
\author[2]{G. Li}
\author[2]{K. T. Yamada}
\author[3]{A. Kirilyuk}
\author[2]{A. V. Kimel}
\author[2]{Th. Rasing}
\author[1]{B. Dieny}
\author[1]{R. C. Sousa}
\author[1]{I. L. Prejbeanu}
\author[1]{L. D. Buda-Prejbeanu}
\affil[1]{Univ. Grenoble Alpes, CEA, CNRS, Grenoble INP, IRIG-SPINTEC, 38000 Grenoble, France}
\affil[2]{Radboud University, Institute for Molecules and Materials, Heyendaalseweg 135, 6525, AJ Nijmegen, The Netherlands}
\affil[3]{FELIX Laboratory, Radboud University, 7 Toernooiveld, 6525, ED Nijmegen, The Netherlands}

\affil[*]{lavilesf@cab.cnea.gov.ar}

\begin{abstract}
Since the first experimental observation of all-optical switching phenomena, intensive research has been focused on finding suitable magnetic systems that can be integrated as storage elements within spintronic devices and whose magnetization can be controlled through ultra-short single laser pulses. We report here atomistic spin simulations of all-optical switching in multilayered structures alternating \emph{n} monolayers of Tb and \emph{m} monolayers of Co. By using a two temperature model, we numerically calculate the thermal variation of the magnetization of each sublattice as well as the magnetization dynamics of [Tb$_n$/Co$_m$] multilayers upon incidence of a single laser pulse. In particular, the condition to observe thermally-induced magnetization switching is investigated upon varying systematically both the composition of the sample (\emph{n,m}) and the laser fluence. The samples with one monolayer of Tb as [Tb$_1$/Co$_2$] and [Tb$_1$/Co$_3$] are showing thermally induced magnetization switching above a fluence threshold. The reversal  mechanism is mediated by the residual magnetization of the Tb lattice while the Co is fully demagnetized in agreement with the models developed for ferrimagnetic alloys. The switching is however not fully deterministic but the error rate can be tuned by the damping parameter. Increasing the number of monolayers the switching becomes completely stochastic. The intermixing at the Tb/Co interfaces appears to be a promising way to reduce the stochasticity. These results predict for the first time the possibility of TIMS in [Tb/Co] multilayers and suggest the occurrence of sub-picosecond magnetization reversal using single laser pulses.
\end{abstract}
\begin{document}

\flushbottom
\maketitle
%
%
\thispagestyle{empty}


\section*{Introduction}

All-optical switching (AOS) phenomena in magnetic materials have been object of thorough research over the last decade due to its fundamental interest and relevance as a possible new write scheme in spintronic devices\cite{Stanciu2007, Kimel2014}. This phenomenon, which was experimentally discovered for the first time in a rare earth (RE)-transition metal (TM) alloy\cite{Stanciu2007}, has then been observed in different multilayers and alloys\cite{Mangin2014,Ghazaly2019,Beens2019}. Two possible mechanisms for the magnetization reversal via ultrashort laser pulses were reported: a helicity-dependent one and another that is thermally-driven. In the helicity-dependent all-optical switching (HD-AOS), the asymmetry of the absorption coming from the magnetic circular dichroism is at the origin of the magnetization switching\cite{Khorsand2012}. Additional mechanisms such as inverse Faraday effect have also been proposed to explain the features of HD-AOS \cite{Stanciu2007, Kimel2005}. On the other hand, the thermally induced magnetic switching (TIMS) generally observed in ferrimagnetic materials, is a process mediated by angular momentum transfer between magnetization sublattices during the remagnetization process, following the incidence of a single laser pulse. From these two mechanisms, HD-AOS is the most commonly encountered and it has been observed in a wide variety of materials. For instance, systematic experimental studies on the switching ability of RE-based and RE-free multilayered systems performed by El-Hadri $et$ $al$.\cite{Salah2017}, demonstrated that engineered synthetic ferrimagnets such as [Ho/Co], [Tb/Co] and [Co/Pt] based multilayers only show HD-AOS. Among all systems switched by HD-AOS, only a few multilayered systems exhibit the possibility of single-shot switching: Pt/Co/Pt/Co/GdFeCo \cite{Gorchon2017}, Pt/Co/Gd \cite{Lalieu2017}, FeCoB/Ta/[Tb/Co]$_\textrm{N}$\cite{Aviles2019, Aviles2020} and Pt/Co/Pt \cite{Vomir2017}. From a theoretical point of view, only two reports discuss the possibility of TIMS in Tb$_x$Co$_{1-x}$ alloys\cite{Moreno2017, Chen2017}. The report from Moreno $et$ $al.$\cite{Moreno2017} based on atomistic spin dynamics modelling concludes that TIMS is possible in Tb$_x$Co$_{1-x}$ \cite{Moreno2017} alloys with compensation temperature higher than room temperature. However previous experiments only reported helicity-dependent AOS with multiple pulses or pure thermal demagnetization\cite{Alebrand2012, Alebrand2014, 2018ZhangARX}. In this framework, it is important to mention that a recent observation of deterministic TIMS in [Tb/Co]-based multilayers evidences a limited fluence range for reversal in [Tb/Co] multilayers systems \cite{Aviles2019}.

\section*{Model}

Our objective is to model synthetic ferrimagnets composed of alternating ferromagnetic layers of Co and Tb with antiferromagnetic exchange coupling between them. The general structure of the multilayered sample is labeled as [Tb$_n$/Co$_m$], where the index $n$ and $m$ correspond to the number of monolayers of each element. In order to explore the probability of TIMS in these multilayers, an atomistic spin solver was used coupled with a two-temperature model (2TM) accounting for the laser heating. Our solver has been developed and benchmarked with respect to previous publications addressing the static magnetic properties and the magnetization dynamics of RE-TM-based systems, such as Tb$_x$Co$_{1-x}$ alloys\cite{Moreno2017} or GdFeCo alloys\cite{Ostler2012}. An array of localized magnetic moments on both Tb and Co sites is considered. The associated Heisenberg Hamiltonian reads: 
 \begin{equation}
\hat{\cal H}=-\frac{1}{2} \sum_{i\neq j} J_{ij} \mathbf{S_\textit{\textrm{i}}} \cdot \mathbf{S_\textit{\textrm{j}}}- \sum_{i} d_{i,z} S_{i,z}^2,
\end{equation}
with $\mathbf{S_\textit{\textrm{i}}}$ the classical spin vectors with unit length associated to the magnetic moment direction on site $i$. The first term represents the Heisenberg exchange energy where the exchange interaction is either between spins of the Tb layer, spins of the Co layer or across the interface between Tb and Co spins. This term therefore contains three interactions $J_{\textrm{Tb-Tb}}$, $J_{\textrm{Co-Co}}$ and $J_{\textrm{Tb-Co}}$, restricted in our model to the nearest neighbors. The second term represents the uniaxial anisotropy with the $d_{i,z}$ constant per atom. The dynamics of the spins is governed by the stochastic Landau-Lifshitz-Gilbert equation of motion: 
\begin{equation}
(1+\lambda_i^2)\mu_i\frac{d\mathbf{S_\textit{\textrm{i}}}}{dt}=-\gamma_i\mathbf{S_\textit{\textrm{i}}} \times [\mathbf{H_\textit{\textrm{i}}}(t)+\lambda_i \mathbf{S_\textit{\textrm{i}}}  \times \mathbf{H_\textit{\textrm{i}}}(t)], 
\end{equation}
where $\gamma_i$,  $\mu_i$ and $\lambda_i$ are the gyromagnetic ratio, the value of the magnetic moment and the damping constant of the spin $i$ respectively. The effective field at each site $i$ is derived from the Hamiltonian $\mathbf{H_\textit{\textrm{i}}}(t)=-\frac{\delta \hat{\cal H}}{\delta \mathbf{S_\textit{\textrm{i}}}}+\zeta_i$ and includes a stochastic thermal contribution $\zeta_i$, accounting for the coupling to the heat bath with the following white-noise properties: 
\begin{equation}
\langle \zeta_i \rangle = 0, \langle \zeta_{i \eta} (0) \zeta_{j \theta}(t)\rangle = 2\delta_{ij}\delta_{\eta\theta}\delta (t) \frac{\lambda_i k_B T_e \mu_i}{\gamma_i},
\end{equation}
where $i$, $j$ correspond to the lattice sites and $\eta$, $\theta$ to the Cartesian components respectively. $T_e$ is the electron temperature, which is related with the laser power during the excitation of a femtosecond long single laser pulse according to a standard 2TM expressed by two coupled equations: 
\begin{equation}
C_e \frac{dT_e(t)}{dt}=-G[T_e(t)-T_{ph}(t)]+P(t)-\frac{C_e}{t_0}[T_e(t)-T_0(t)]
\label{2tm1}
\end{equation}
\begin{equation}
C_{ph} \frac{dT_{ph}(t)}{dt}=G[T_e(t)-T_{ph}(t)]-\frac{C_{ph}}{t_0}[T_{ph}(t)-T_0(t)],
\label{2tm2}
\end{equation}
where $T_{ph}(t)$, $C_e$ = $\gamma T_e$ and $C_{ph}$  correspond to the temperature of the phonon bath, the heat capacity of electrons and the heat capacity of the phonons respectively. $G$ is an electron-phonon coupling parameter and $P(t)$ is the external power injected by the laser pulses defined as $P(t) = (I_0 F) e^{-(t/\tau_p)^2}$, where $I_0$, $F$ and $\tau_p$ correspond respectively to the laser energy absorbed by the system, the fluence and the laser pulse duration. Eqs.~\ref{2tm1} and \ref{2tm2} also include a Newton's cooling term which drives the system back to room temperature after the incidence of the laser pulse \cite{Anisimov1974,2018Gridnev,2019Gridnev}. According to Newton's law the rate of heat loss is proportional to the temperature difference between the body and its surroundings and involves a characteristic time $t_0$. It is important to mention that the heat produced by the laser pulse will slowly diffuse away from the electron system towards the heat sink fixed at room temperature after a few ps. In experiments, this is the required time for heat to dissipate to the substrate\cite{Cornelissen2016}.

We have considered a multilayered system consisting of 30$\times$30$\times$30 fcc cells with periodic boundary conditions. We assumed that there are no size differences between RE and TM atoms. Only exchange with the 12 nearest neighbors is taken into account. A schematic of the atom distribution is shown in Fig. S1 in the Supplementary data file. The local uniaxial anisotropy, the effective magnetic moment of Co and Tb sub-lattices and the interatomic exchange interactions values were taken from Moreno $et$ $al$., which reports atomistic calculation of TIMS in Tb$_x$Co$_{1-x}$ alloys. The values of various parameters are listed in Table~\ref{tabla}. The magnetic moment as well as the intra-lattice exchange correspond to the bulk value of pure Co respectively Tb. Normally, it is expected that these parameters are sample dependent since the precise composition (number of multilayers, defects), the nature of the substrate, material deposition technique as well as annealing condition may influence the precise arrangement of magnetic moments in the lattice. We made the choice to keep these parameters constant upon varying the composition of the multilayers for comparison with a large panel of compositions. Calibration with respect to the real samples combined with ab-initio computation are desirable for a complete quantitative analysis but here we prefer to consider a perfect model system. The time integration of the above-mentioned equations was performed according to a Heun scheme using a time step below 0.1 fs. The parameters of Newton’s law are only impacting the system behavior at longer timescales typically above 1 ns. 

\begin{table}[htbp]
\begin{center}
\caption{List of parameters used for the numerically calculated static and dynamic properties of Tb$_n$Co$_m$ multilayers.}
\begin{tabular}{cc}
\hline
Parameter & Value \\
\hline \hline
Co magnetic moment ($\mu_{Co}$) & 1.61 $\mu_B$ \\ 
Tb magnetic moment ($\mu_{Tb}$) & 9.34 $\mu_B$ \\ 
Exchange constant Co-Co ($J_{\textrm{Co-Co}}$) & 5.9$\times$10$^{-21}$ J \\
Exchange constant Tb-Co ($J_{\textrm{Tb-Co}}$) & -1.0$\times$10$^{-21}$ J \\
Exchange constant Tb-Tb ($J_{\textrm{Tb-Tb}}$) & 8.2$\times$10$^{-22}$ J \\
Uniaxial anisotropy constant ($d_{i,z}^{Co}$) & 3.73$\times$10$^{-23}$ J \\
Uniaxial anisotropy constant ($d_{i,z}^{Tb}$) & 2.16$\times$10$^{-22}$ J \\ 
Heat capacity of phonons $C_p$ & 3.0$\times$10$^{6}$ Jm$^{-3}$K$^{-2}$\\
Coupling parameter $G$ & 17.0$\times$10$^{17}$ Jm$^{-3}$K$^{-1}$\\
Heat capacity of electrons $C_e$ (at RT) & 2.1 $\times$10$^{5}$ Jm$^{-3}$K$^{-1}$\\
Characteristic time $t_0$ & 10 ps\\
Damping parameter $\lambda$ & 0.03 \& 0.05\\
Laser intensity $I_0$  & 3.0$\times$10$^{-19}$ m$^{-1}$s$^{-1}$\\
\hline
\end{tabular}
\label{tabla}
\end{center}
\end{table}

\section*{Static magnetic properties of [$\textrm{Tb}/\textrm{Co}$] multilayers}

\begin{figure}[ht]
\includegraphics[width=16cm]{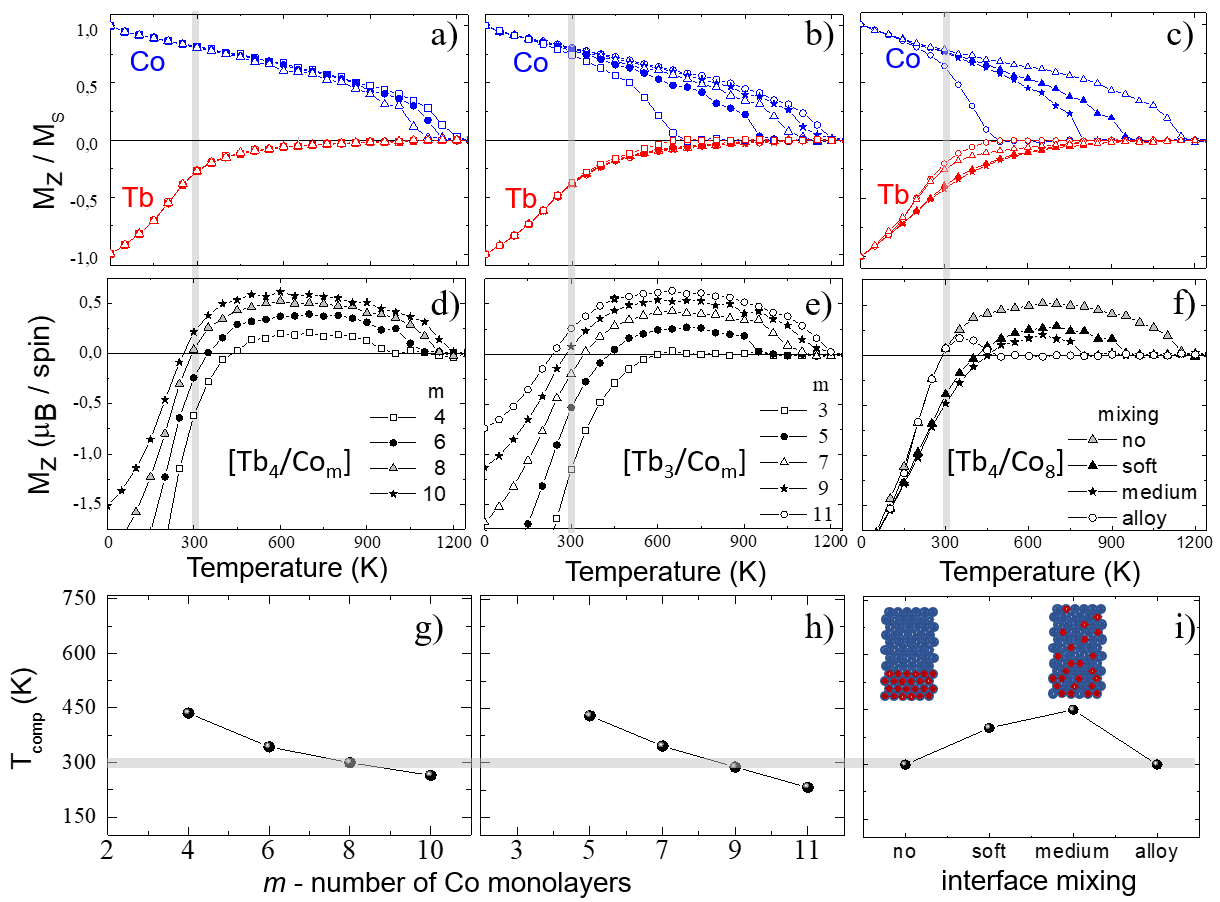}
\caption{\label{fig1} Thermal variation of magnetization of [Tb$_n$/Co$_m$] multilayers. The subscripts indicate the number of atomic layers. (a), (b), (c) Normalized thermal variation of the magnetization of Tb and Co sublattices. (d), (e), (f) Reduced magnetic moment of sample [Tb$_n$/Co$_m$]. Compensation temperature as a function of the number of Co monolayers (g), (h) and intermixing (i). The insets from (i) show the composition of the sample with perfect interfaces and medium intermixing respectively. The grey lines indicate the room temperature.}
\end{figure}

The first aim of the study was to obtain the equilibrium properties of the multilayer as a function of their composition. Thus, the equilibrium magnetization was calculated as the spatial and time average of the magnetic moments over the volume of the sample. Figure~\ref{fig1} summarizes the results for three series of samples [Tb$_4$/Co$_m$] (Fig.~\ref{fig1}a, d, g), [Tb$_3$/Co$_m$] (Fig~\ref{fig1}b, e, h) and [Tb$_4$/Co$_8$] with various degree of intermixing of the multilayered sample (Fig~\ref{fig1}c, f, i). The normalized $M_s(T)$ curves for each individual Tb and Co sub-lattice (Fig~\ref{fig1}a, b, c), show typical decrease versus temperature due to increasing spin fluctuations.

In the case of ideal multilayered samples (Fig.~\ref{fig1}a, b), the magnetization of the Co sublattice closely follows the typical behavior of a ferromagnet since the intra-layer exchange interaction is strong. The Curie temperature of the coupled multilayers is around or above 1000 K approaching the Curie temperature of the Co sub-lattice. The Tb sub-lattice having a weaker intra-layer exchange interaction exhibits magnetic ordering above its own bulk Curie temperature because of the interaction with the Co sublattice, the decrease of its magnetization follows an almost linear dependence with  the temperature. The overall magnetization curves (Fig.~\ref{fig1}d, e) indicate that the multilayers behave as synthetic ferrimagnets with a corresponding magnetic compensation point (T$_{comp}$). An exception is found for the sample [Tb$_3$/Co$_3$] (m=n=3) having also much smaller Curie temperature around 705 K. Our simulations also confirm that the compensation temperature decreases almost linearly with the Co total magnetic moment as can be seen in Fig.~\ref{fig1}g, h. Two samples [Tb$_4$/Co$_8$]  and [Tb$_3$/Co$_9$]  have a compensation point around room temperature and thus are suitable for TIMS experiments. 

The impact of the interfacial intermixing has been investigated in the case of the [Tb$_4$/Co$_8$] sample. For that purpose, a soft, medium and strong intermixing were compared keeping constant the number of Co and Tb atoms. A moderate intermixing shifts the compensation point upwards, but also progressively reduces the Curie temperature. In the limit case of an homogeneous alloy, the Curie temperature is drastically reduced since the Co magnetic moments have less Co neighbors.

\section*{Dynamic magnetic properties of [$\textrm{Tb}/\textrm{Co}$] multilayers}

\begin{figure}[ht]
\begin{tabular}{c}
\includegraphics[width=8.5cm]{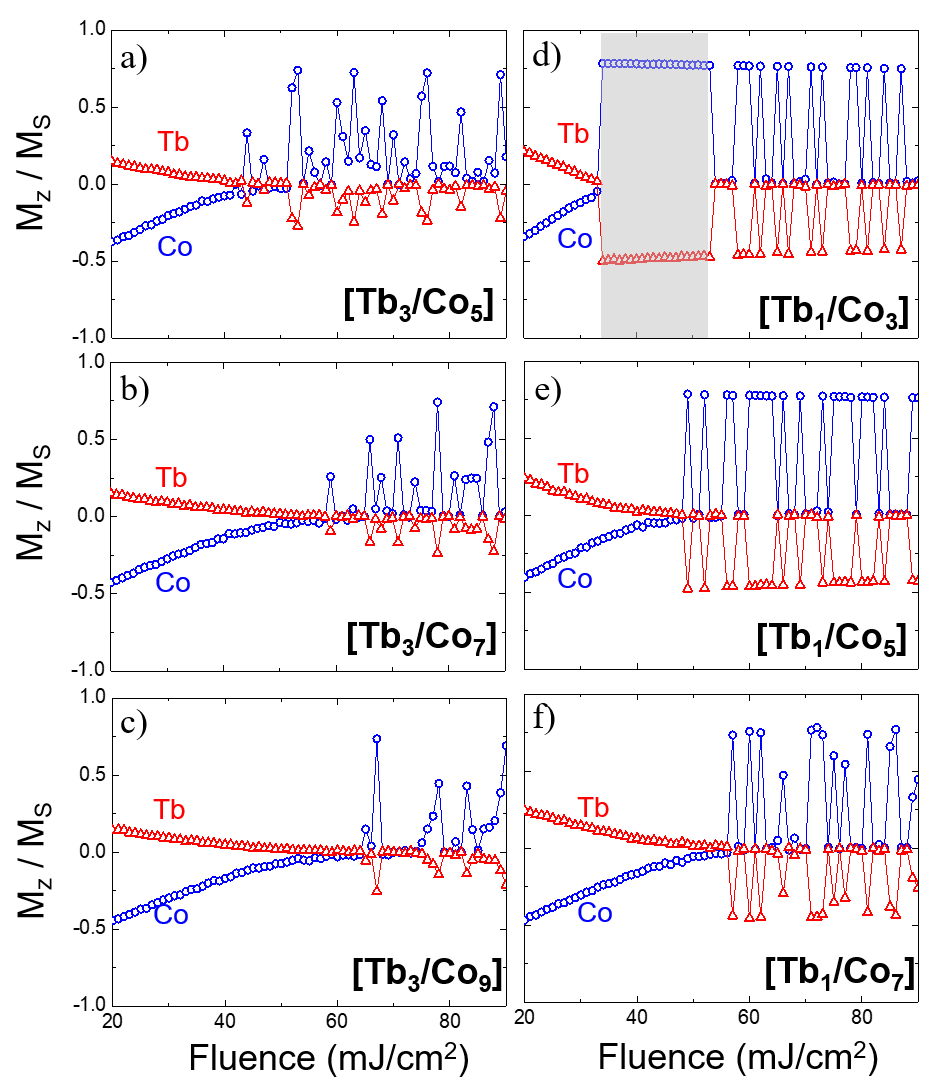}\\
\end{tabular}
\caption{\label{fig2} Normalized magnetization of Tb and Co sublattices as a function of the laser fluence. $M_z/M_s$ ratios are extracted from the magnetization curves of the [Tb$_n$/Co$_m$] multilayers 20 ps after the single laser pulse has interacted with the system. Each peak indicates switching, where Tb goes to a negative magnetization and Co goes to a positive magnetization. The noise observed in these graphs reflects the impact of random thermal fluctuations on the switching dynamics. A deterministic switching window is marked with light grey color for [Tb$_1$/Co$_3$] multilayers.}
\end{figure}

The switching properties of the [Tb$_n$/Co$_m$] multilayers after the incidence of a single laser pulse were analyzed using the two temperature model (Eq. 4 and 5). The total duration of the laser pulse is set at 50 fs and its fluence is varied up to 120 mJ/cm$^2$. The magnetization of each sub-lattice was recorded at each time step and the value at 20 ps after illumination is included in the diagrams presented in Fig.~\ref{fig2} for two series of samples [Tb$_3$/Co$_m$] and [Tb$_1$Co$_m$] respectively.

Figs.~\ref{fig2}a-c, corresponding to [Tb$_3/$Co$_m$], shows random peaks of $M_z/M_s$ after 20 ps, indicating that the magnetization reversal of each sub-lattice occurs randomly for $n$= 3. However, the multilayer [Tb$_1$/Co$_3$] in Fig.~\ref{fig2}d presents a clear deterministic switching window for fluence values between 35 and 50 mJ/cm$^2$ in which magnetization reversal occurs. For this composition, the exchange interaction between Tb and Co dominates the switching process. TIMS have been observed for [Tb$_1$/Co$_3$] and for [Tb$_1$/Co$_2$] with our model. However, Figs.~\ref{fig2}d-f evidence that the fluence switching window rapidly disappears for $m \geq$ 5. This is a strong indication that the magnetization switching in the [Tb$_n$/Co$_m$] requires a non-negligible number of interfaces with antiferromagnetic exchange in order to fully reverse the magnetization after each single laser pulse. As a matter of fact, the high number of antiferromagnetic interfaces present in [Tb$_1$/Co$_3$] system supports the theory according to which the switching originates at one interface and is followed by a propagation of the switching front towards the inner part of the layers comprising multilayer until a complete switching is achieved. The fluence threshold to trigger the switching increases with the number of Co monolayers, which is consistent with the increase of the Curie temperature. On the other hand, it is also important to notice that [Tb$_1$/Co$_2$] and [Tb$_1$/Co$_3$], whose Curie temperature and compensation temperature (around 750 K) are much higher than room temperature are the only systems that showed a clear fluence window where TIMS can be observed with our model. These results are in agreement with the observation of Moreno \textit{et al}. \cite{Moreno2017}, in Tb$_x$Co$_{1-x}$ alloys, in which systems with a compensation temperature higher than room temperature undergo thermally-induced magnetization switching in a wider fluence window \cite{Moreno2017}. 

There is also a remarkable difference between the magnetization dynamics behavior of the sublattices in the [Tb$_1$/Co$_3$] and the [Tb$_3$/Co$_9$] systems as a function of the laser fluence shown in Figs.~\ref{fig2}c and~\ref{fig2}d. Although, these two systems have the same proportion of Tb and Co thicknesses, the switching window was fully suppressed, and additionally the response of the magnetization upon photo-excitation requires at least 60 mJ/cm$^2$ to induce partial switching. This observation highlights and confirms the key role of the interface with antiferromagnetic exchange coupling in the switching mechanism of multilayer structures\cite{Xu2016}.

\begin{figure}[ht]
\begin{tabular}{c}
\includegraphics[width=17cm]{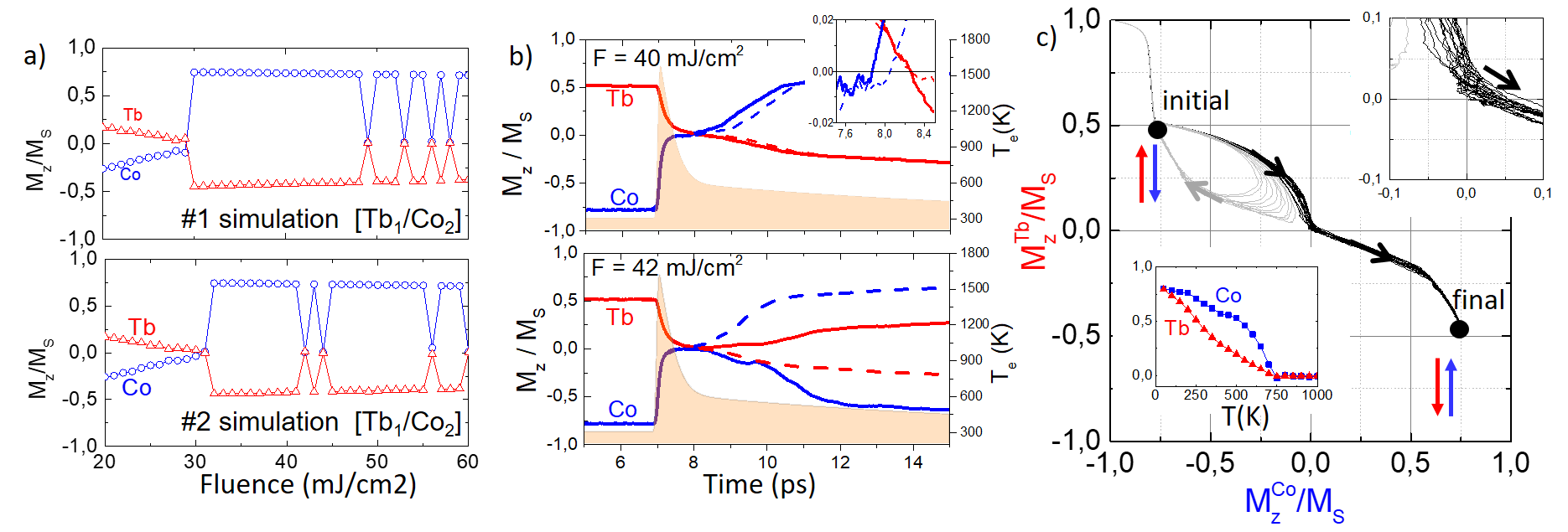}\\
\end{tabular}
\caption{\label{fig3} a) Normalized magnetization of Tb and Co as a function of the laser fluence. $M_z/M_s$ ratios were extracted from the magnetization curves of two different simulations of the [Tb$_1$/Co$_2$] system. b) Time evolution of the magnetization of Tb and Co sublattices, M$^\textrm{Co}$ (blue) and M$^\textrm{Tb}$ (red), after the application of a single laser pulse at $t$ = 7 ps in the system [Tb$_1$/Co$_2$] for two different fluences: 40 mJ/cm$^2$ (top) and 42 mJ/cm$^2$ (bottom). The two different events for each fluence value are represented by the solid and dashed lines. Inset: $M_Z$/$M_S$ during 7.6 $<t<$ 8.4 ps. Evolution of the electronic temperature after the single laser pulse interacts with the system at $t$ = 7 ps (pink background). The single laser pulse duration is 50 fs. c) Magnetization of Tb and Co sublattices in a magnetization phase map. Upper inset: Pathways followed by the magnetizations in a vicinity of zero. The magnetic moment of the Co sublattice always crosses zero before the magnetic moment of the Tb sublattice. Lower inset: Magnetization of the sublattices as a function of the temperature. The effective Curie temperature of the system is close to 750 K.}
\end{figure}

Once the fluence switching window was obtained, further magnetization dynamic simulations were carried out in order to check whether thermally-induced magnetization switching can be fully deterministic. Although we have obtained a clear fluence switching window for [Tb$_1$/Co$_3$] and for [Tb$_1$/Co$_2$], additional simulated events showed that for some values between 35 and 50 mJ/cm$^2$, thermally-induced magnetization switching is not fully deterministic. Fig.~\ref{fig3}a shows the normalized magnetization of Tb and Co as a function of the laser fluence for [Tb$_1$/Co$_2$], between 30 - 50 mJ/cm$^2$ obtained from two different experiments. It can be seen that for fluences 42 mJ/cm$^2$ and 44 mJ/cm$^2$ the switching is not deterministic. Fig.~\ref{fig3}b plots the temporal dependence of the magnetization of the two different events for a fluence F = 40 mJ/cm$^2$ and F = 42 mJ/cm$^2$. The two events at each fluence value are indicated by the solid and dashed lines. On the magnetization curves in Fig.~\ref{fig3}b, the single laser pulses were sent at time $t$= 7 ps. During the first picosecond after the heating, the single laser pulse generates a strong non-equilibrium state in the Tb and Co sublattices originating the decoupling of the sublattices and a subsequent demagnetization at different rates. It is observed in the inset of Fig.~\ref{fig3}b that the magnetization of the Co sublattice crosses zero before the one of the Tb sublattice, due to the demagnetization of Co sublattice is faster than that of the Tb sublattice. The total magnetization of the system, including the transient ferromagnetic state created during the next picosecond is mediated by the antiferromagnetic exchange interaction between the Tb and Co spins. Due to this interaction and for a small window of time ($t=$ 7.8 – 8.2 ps), the magnetic moments of the Co sublattice grow and align parallel with respect to the magnetic moments of the Tb sublattice. Finally, Tb magnetic moments will also cross zero and the remagnetization process ($t >$ 8.2 ps) will reorient the total magnetization in an antiparallel magnetic polarity with respect to the initial state. The transient ferromagnetic state, is characteristic of a thermally induced magnetization switching as was already discussed by Radu \textit{et al}. \cite{Radu2011} and now also confirmed by our simulations. As can be seen in the right part of plots in Figure~\ref{fig3}b, the electronic temperature increases up to 1600 K and rapidly decreases close to 500 K before $t$ = 8 ps. The temporal evolution of the electronic temperature during the simulations indicates that the maximum temperature (1600 K) of the electronic bath is higher than the Curie temperature of the [Tb$_1$/Co$_2$] multilayer (bottom inset of Fig.~\ref{fig3}c). 

In order to visualize the evolution of the magnetization during the ferromagnetic transient regime, we plot the magnetic moment of the Tb and Co sublattices in a magnetization phase diagram (Fig.~\ref{fig3}c) based on the model proposed by C. Davies \textit{et al}. \cite{Davies2020, Mentink2012}. The top inset of Figure \ref{fig3}c zooms in the trajectories followed by the Tb and Co magnetic moments during the demagnetization and reorientation processes within the third quadrant of the diagram phase (corresponding to the ferromagnetic transient state). In accordance with the duration of the transient ferromagnetic state extracted from the inset of Fig ~\ref{fig3}b (0.4 ps), the pathways that follow the magnetic moments of the sublattices are contained within the values 0 $<$ M$^\textrm{Co}_z$/M$_s<$ 0.05 and 0 $<$ M$^\textrm{Tb}_z$/M$_s<$ 0.05 for a very short period of time. This also indicates that the Tb sublattices only posses a small magnetic moment value to transfer to the Co sublattice once the Co magnetic moment reaches zero. On the other hand, comparing the duration of the transient ferromagnetic state of our [Tb$_1$/Co$_2$]: 0.4 ps, with the one experimentally observed in GdFeCo layers that is around 1 ps \cite{Radu2011}, or with the one extracted from numerical simulations in CoGd-based systems, also around 1 ps, it might be  expected an increase of the switching probability in the [Tb/Co]-based system by increasing the duration of the transient ferromagnetic state. Experimentally, this might be achieved by altering the demagnetization rate though the tunning of the magnetic anisotropy during the growth of the Co and Tb layers. 

\begin{figure}[ht]
\includegraphics[width=8.5cm]{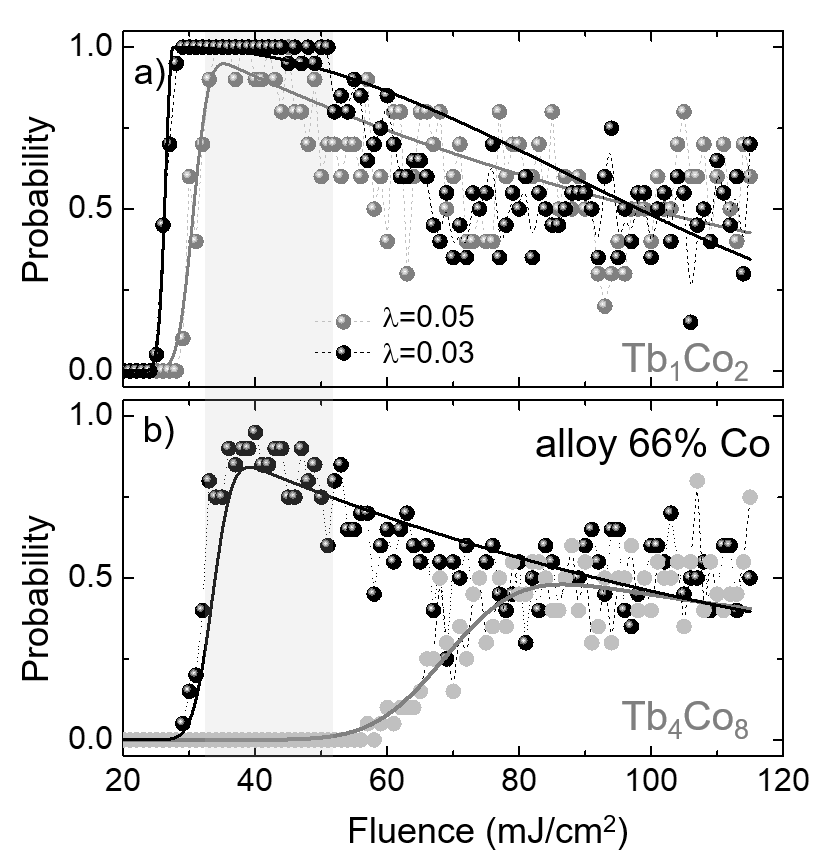}\\
\caption{\label{fig4} a) Switching probability as a function of the laser fluence for the system [Tb$_1$/Co$_2$] over 20 events with two different values of damping constant $\lambda$ = 0.03 and 0.05. There is a maximum of the switching probability for F = 35 mJ/cm$^2$ for the case of $\lambda$ = 0.05. b) Switching probability as a function of the laser fluence for the system [Tb$_4$/Co$_8$] and for a homogeneous TbCo alloy with 66 $\%$ of Co. Black and grey solid lines are a guide to the eyes.} 
\end{figure}

As shown above, a fully reversal of the magnetization is not always observed for repeated attempts at a given fluence. Fig.~\ref{fig3}b showed the magnetization dynamics upon the incidence of a single laser pulse with a fluence F = 42 mJ/cm$^2$, in which the final state of the magnetization of the Co and Tb sublattices does not show switching (dashed lines) after 20 ps. To further investigate the stochasticity of the magnetization reversal, we calculated the switching probability in the [Tb$_1$/Co$_2$] multilayer by repeating 20 times the simulation of the magnetization dynamics. Fig.~\ref{fig4}a illustrates the switching probability for the [Tb$_1$/Co$_2$] system, which has a maximum probability at 35 mJ/cm$^2$. It can also be observed that for the fluence window between 35 and 50 mJ/cm$^2$, indicated by the gray rectangle the switching probability is above 80 \%. For fluence values above this reversal window, the stochasticity increases, while the switching probability rapidly decreases for higher fluence values. We also explored the influence  of the damping constant value ($\lambda$) in the switching probability of the [Tb$_1$/Co$_2$] multilayer. By decreasing the damping constant from $\lambda$ = 0.05 to $\lambda$ = 0.03 in the Landau-Lifshitz-Gilbert equation, we were able to adjust the switching probability and the fluence window of the [Tb$_1$/Co$_2$] system. As can be seen in Fig.~\ref{fig4}a, the switching probability as a function of the laser fluence over 20 events increased to 100 \% for a fluence window between 30 and 50 mJ/cm$^2$ for $\lambda$ = 0.03. The decrease of the damping constant impacts the remagnetization time increasing the switching probability of the [Tb$_1$/Co$_2$] system. Although, we have used the same damping constant value for Tb and Co atoms to determine the switching probability, as was done in similar works \cite{Radu2011, Moreno2017}, the exploration of element-specific damping could provide more information about the switching dynamics of the system as suggested by recent reports \cite{Ceballos2019}. Our simulation and experiments\cite{Aviles2020} were performed using fs-long pulses to evaluate the optical switching in [Tb/Co]-based systems and only [Tb$_1$/Co$_2$] and [Tb$_1$/Co$_3$] showed a clear fluence window for the observation of optical switching using 50 fs pulses (Fig. S2 in the Supplementary data file). The possibility to switch the magnetization in these systems with ps-long laser pulses using atomistic simulations remains unexplored.

As expected, the switching probability significantly drops for thicker Tb and Co layers, and similarly to the case of [Tb$_1$/Co$_3$] and [Tb$_3$/Co$_9$] discussed before, the magnetization dynamics of the [Tb$_1$/Co$_2$] and [Tb$_4$/Co$_8$] multilayers, which have the same proportion of Tb and Co thickness, also presents completely different behaviors. The switching probability is below 60 \% and the minimum fluence value to excite or induce magnetization reversal is around 60 mJ/cm$^2$, as can be seen in Fig~\ref{fig4}b. The section S3 of the Supplementary information file shows the normalized magnetization of Tb and Co as a function of the laser fluence and also the time evolution of both magnetization after the laser heats the sample. Contrary to the switching of the [Tb$_1$/Co$_2$] system, the results obtained in the multilayer [Tb$_4$/Co$_8$] showed that when there is no controlled magnetization switching, as a result of the noise introduced by the thermal fluctuations in the two-temperature model, the construction of the trajectories in the magnetization phase diagram does not follow the model suggested by Davies \textit{et al}. \cite{Davies2020, Mentink2012} (see Fig. S3c in the supplementary data file).

In order to assess the hypothesis that the antiferromagnetic exchange coupling is relevant to the TIMS process, we investigated the influence of the intermixing in the alloy structure and calculated the switching probability in a TbCo alloy system with 66 \% of Co. Fig. S4 of the supplementary data file shows the distribution of the Co and Tb atoms for the ideal sample [Tb$_4$/Co$_8$] as well as that of the samples affected by the intermixing at the interfaces. Fig~\ref{fig4}b shows that the dependence of the thermally-induced magnetization switching probability as a function of the fluence is comparable to that of the [Tb$_1$/Co$_2$] multilayer system. The study performed in the alloy with 66 \% of Co in the TbCo alloy can also be compared with the one of 68 \% of Co in the TbCo alloy reported in Ref. 11, for which a narrow switching window is expected at a much lower pulsed laser fluence. On the other hand, the enhancement of the probability in the alloy case is also in agreement with other reports highlighting the role of intermixing in synthetic ferrimagnets \cite{Beens2019a}.

The results presented in this work indicate that not only thermal fluctuations but also the values of the damping constant and the intermixing are critical and expected to play an important role in the successful optical switching of ultra-thin [Tb$_n$/Co$_m$]. Further studies will be needed to understand the influence of these parameters in detail. On the other hand, signatures of stochastic switching have been found in preliminary electrical test of magnetic junctions with Tb/Co-based multilayer electrodes (Sec. S5 of the supplementary data file), therefore additional experiments that fully support our simulations need to be performed. 

\section*{Conclusion}
In conclusion, we applied an atomistic spin model to [Tb/Co] multilayered structures and calculated the thermal variation of its magnetization and the magnetization dynamics upon incidence of single laser pulses. After introducing the physical influence of the laser pulse into the model, the all-optical switching of [Tb/Co] multilayers was simulated. We found that [Tb$_1$/Co$_2$] and [Tb$_1$/Co$_3$] present a fluence window between 35 and 50 mJ/cm$^2$ in which the multilayers can be switched by a single laser pulse, presenting an error rate below 5 \%. Stochasticity of the switching process was determined by the calculation of the probability over 20 switching attempts. We also pointed out the role of the damping constant value and the intermixing in the switching occurrence. Lower values of damping constant and intermixed interfaces enhance the switching probability in our system. These results predict for the first time the possibility of TIMS in ultra-thin [Tb/Co] multilayers by using atomistic spin simulations and suggest the occurrence of sub-picosecond magnetization reversal using single laser pulses in a [Tb/Co]-based multilayer structures. The deterministic switching and the estimation of the required fluence are in agreement with recent experimental observations of single-shot all-optical switching in [Tb/Co]$_5$-based multilayers\cite{Aviles2019, Aviles2020}. The prediction of single-shot switching in [Tb/Co] multilayers based on numerical simulation results and its confirmation by experimental observations, show that this type of simulations can be used as a guiding tool to find appropriate systems exhibiting single pulse all-optical switching. The physical mechanisms underlying thermally induced magnetization switching in [Tb$_1$/Co$_2$] is in agreement with the model proposed by Davies \textit{et al}. \cite{Davies2020,Mentink2012} Furthermore, in thicker layers the demagnetizing processes dominate, the magnetization of the two ferromagnetic layers is divided in small magnetic domains preventing thus the complete reversal.



\bibliography{sample}



\section*{Acknowledgements}
This research has received funding from the European Union’s Horizon 2020 research and innovation program under FET-Open Grant Agreement No. 713481 (SPICE). 
\section*{Author contributions statement}
Z.J. and L.D.B.P developed the numerical solver, L.F., Z.J. and L.A.G. performed the numerical simulations; L.A.F., L.F., I.L.P., L.D.B.P., B.D. and R.C.S. discussed and analyzed the numerical results; L.A.F., G. L and K. Y performed the electrical measurements; L.A.F., G. L., K. Y., A. K., A. V. K., Th. R., I.L.P., R.C.S. and L.D.B.P. discussed and analyzed the electrical measurements, L.A.F. wrote the manuscript with contributions from I.L.P, R.C.S, Th. R, B.D. and L.D.B.P.; I.L.P., R.C.S. B.D. and L.D.B.P. supervised the project.

\section*{Additional information}
$^+$Current address: Centro At\'omico Bariloche, Comisi\'on Nacional de Energ\'ia At\'omica (CNEA), Consejo Nacional de Investigaciones Cient\'ificas y T\'ecnicas (CONICET), Avenida Bustillo 9500,  Bariloche, Rio Negro, Argentina

\section*{Competing interests}
The authors declare no competing interests.

\end{document}